# Does Probabilistic Constellation Shaping Benefit IM-DD Systems without Optical Amplifiers?

Di Che, Junho Cho, and Xi Chen

*Abstract*—Probabilistic constellation shaping (PCS) has been widely applied to amplified coherent optical transmissions owing to its shaping gain over the uniform signaling and fine-grained rate adaptation to the underlying fiber channel condition. These merits stimulate the study of applying PCS to short-reach applications dominated by intensity modulation (IM) – direct detection (DD) systems. As commercial IM-DD systems typically do not employ optical amplification to save the cost and power consumption, they are no longer subject to an average power constraint (APC) but a peak power constraint (PPC), which poses unique challenges to take full advantages of PCS. This paper provides a comprehensive investigation of PCS in IM-DD systems without optical amplifiers. In particular, we reveal that if the transmitter enhances the peak-to-average power ratio of the signal, a PPC system can be partially or even fully converted to an APC system in which the classical PCS offers its merits. The findings are verified through an IM-DD experiment using 4- and 8-ary pulse amplitude modulations.

*Index Terms*—Optical communication, intensity modulation, direct detection, pulse amplitude modulation, probabilistic constellation shaping, rate adaptation.

## I. INTRODUCTION

PROBABILISTIC constellation shaping (PCS) has achieved remarkable success in *coherent* optical transmissions since its first demonstration in optical fiber transmission simulation [1] and experiment [2] in 2015. PCS shapes the probability of discrete constellation points to approximate the continuous Gaussian signaling. It improves the information rate (IR) over uniform signaling, because: *(i)* Gaussian signaling maximizes the IR of the additive white Gaussian noise (AWGN) channel under an *average power constraint (APC)*, and *(ii)* optical fiber channels can be accurately modeled as AWGN channels if optical amplified spontaneous emission (ASE) is the predominant channel impairment and nonlinear inter-channel and intra-channel interference is regarded as noise [3]-[5]. The increase in IR for the same optical launch power (or the enhanced energy efficiency for the same IR) achieved by PCS [6],[7] compared to uniform signaling is called *shaping gain*. In addition to the shaping gain, PCS offers a significant benefit of fine-grained *rate adaptation*. The PCS can be easily combined with off-the-shelf forward error correction (FEC) codes using the probabilistic amplitude shaping (PAS) architecture [8]. While the IR can be continuously adjusted by both FEC and PCS to match the underlying channel condition, in practice, PCS can achieve rate adaptation in a much simpler way with finer granularity and at a lower implementation cost than the rate adaptation based only on FEC [9].

Owing to the merits of shaping gain and rate adaptation, PCS has been considered for short-reach interconnects dominated by intensity modulation (IM) and direct detection (DD) [10]-[21]. Compared with long-haul coherent systems, IM-DD systems have a few unique characteristics: *(i)* signals are modulated with unipolar intensities instead of complex optical fields, *(ii)* short-reach IM-DD links typically do not use optical amplification, *(iii)* the system is therefore limited by the peak-power constraint (PPC) of transceivers rather than the APC imposed by nonlinear interference (NLI), *(iv)* the dominant system noise comes from the transceiver rather than the optical channel. Although PCS has been applied to IM-DD systems using various types of modulation schemes and probability distributions, there is no systematic analysis on how the above unique characteristics of IM-DD systems influence the performance of PCS. This paper aims to provide a comprehensive analysis of PCS in IM-DD systems *without optical amplifiers* (throughout the paper, the term "IM-DD systems" refers to systems without amplifiers unless otherwise specified). We reveal whether PCS benefits IM-DD systems, and if so, in which system conditions does PCS maximize its benefits. Considering the recent trend of applying coherent techniques to short-reach interconnects like the standardization of 400G-ZR, the findings of this paper may also be extended to future short-reach coherent systems without optical amplifiers limited by PPC and transceiver noise [22-25].

The paper is organized as follows. Section II briefly reviews the benefits of PCS in APC systems and the evaluation metrics of the performance that will be used throughout the paper; Section III presents a mathematical model for IM-DD systems that exhibits clearly distinguished characteristics from coherent systems; Sections IV and V analyze various signal distributions for PCS under different conditions in IM-DD systems; Section VI experimentally verifies the findings of Sections III through V, and Section VII concludes the paper.

## II. PCS IN AWGN CHANNELS

We focus on the pulse amplitude modulation (PAM) in this work, which is a widely used modulation in commercial IM-

The authors are with Nokia Bell Labs, Holmdel, NJ 07733 USA (e-mail: di.che@nokia-bell-labs.com; junho.cho@nokia-bell-labs.com; xi.v.chen@nokia-bell-labs.com). Corresponding author: Di Che.

DD transceivers. Given an $M$-ary PAM (PAM-$M$) alphabet $\mathcal{X}$, the entropy of a symbol $X$ drawn from the alphabet $\mathcal{X}$ with the probability distribution $P_X$ is calculated as

$$H(X) = -\sum_{x \in \mathcal{X}} P_X(x) \log_2 P_X(x). \quad (1)$$

This quantifies the theoretically maximum information content that can be conveyed by the symbol $X$. Under an APC, the distribution that maximizes $H(X)$ is the Maxwell-Boltzmann (MB) distribution [26]

$$P_X(x) = \frac{e^{-\lambda x^2}}{\sum_{x' \in \mathcal{X}} e^{-\lambda x'^2}}, \quad (2)$$

where the single free parameter $\lambda \geq 0$ determines the shape of the distribution, the entropy $H(X)$, and the average symbol energy $\sum_{x \in \mathcal{X}} P_X(x)|x|^2$ to meet the APC. Conversely, the MB distribution minimizes the average energy of the symbol $X$ for the same entropy. In the context of data transmissions, when a target $H(X)$ and an APC are given in an AWGN channel, a transmitter should scale the PAM-$M$ constellation $\mathcal{X}$ by a scaling factor $\Delta$ as $\Delta \mathcal{X}$ such that the symbols $X \in \Delta \mathcal{X}$ with the distribution $P_X$ meet the given APC; such scaling factor $\Delta$ (hence the minimum Euclidean distance between the symbols $X$) is maximized when $P_X$ is an MB distribution. Therefore, the MB distribution is widely used for PCS in the coherent optical communication that is subject to an APC. Under a PPC, however, the distribution which maximizes $H(X)$ on the alphabet $\Delta \mathcal{X}$ is the uniform distribution $P_X(x) = 1/|\mathcal{X}|$, as commonly used by traditional IM-DD systems in the form of uniform PAM-$M$. Apparently, the benefits of different distributions for PCS appear differently depending on the system constraint.

If an AWGN channel is subject to an APC, signal-to-noise ratio (SNR) is the figure of merit to characterize the channel quality. At a fixed SNR, the performance of a soft-decision (SD) FEC code with a fixed code rate $R_c$ can vary with the modulation format, namely, the PAM alphabet size, distribution $P_X$ and entropy $H(X)$. To determine if error-free transmission is accomplished for various modulations at any given SNR, we use normalized generalized mutual information (NGMI) as a performance metric in the paper, which is a modulation-agnostic figure of merit for binary SD-FEC [7]. Namely, if the NGMI measured from an IM-DD system is greater than the NGMI threshold (denoted as $NGMI^*$) of the selected SD-FEC scheme, we can declare an error-free transmission with high confidence, and the system achieves an IR of

$$R = R_s - \underbrace{(1 - R_c) \log_2 M}_{(a)} \quad (3)$$

in bits per PAM symbol, where $R_s$ denotes the shaping rate that quantifies the number of information bits contained in a shaped PAM symbol in the absence of noise and the term (a) quantifies the reduction of IR due to the FEC overhead. An ideal shaping algorithm realizes $R_s = H(X)$ in Eq. (3), but a non-ideal pragmatic shaping algorithm provides a shaping rate of $R_s < H(X)$, inducing a rate loss of $H(X) - R_s$ bits per PAM symbol. The NGMI is an SD analogue of the Q factor for HD FEC (indeed, the Q factor is a special type of the NGMI). An ideal FEC coding scheme has $NGMI^* = R_c$, but a non-ideal pragmatic FEC coding scheme has $NGMI^* > R_c$, with a rate loss of $NGMI^* - R_c$.

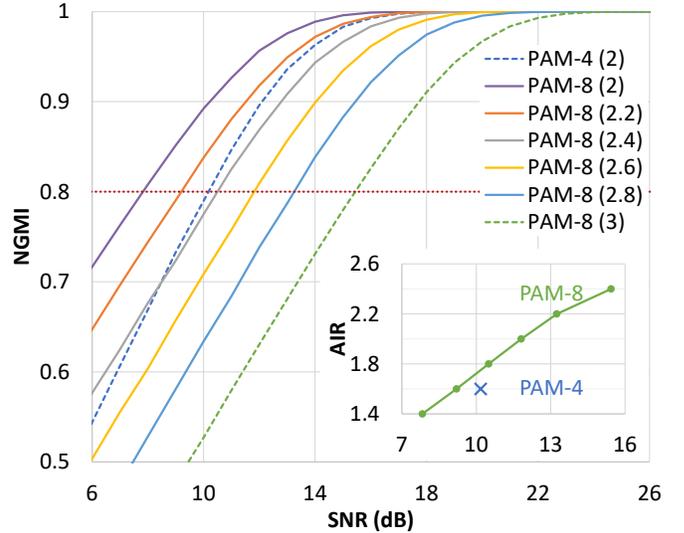

Fig. 1. NGMI of PCS PAM signals in AWGN channels with various SNR. In the legend, the numbers in parentheses are the source entropy in bit/symbol. Inset: rate adaptation assuming an ideal fixed-rate SD-FEC ($NGMI^* = 0.8$).

In this paper, we study the benefits of PCS by adjusting the IR solely by shaping the symbol distribution at a fixed FEC rate, since rate adaptation by FEC is considered too costly for short-reach IM-DD systems [9]. Also, since this paper aims to clarify the benefits of PCS in various IM-DD system realizations, we assume ideal rate-adaptable shaping with $R_s = H(X)$ and ideal rate-0.8 FEC coding with $NGMI^* = R_c = 0.8$. The IR obtained in this coded modulation setup represents an upper bound of the IR that can practically be achieved under the code rate constraint of $R_c = 0.8$, hence called an achievable IR (AIR). The analysis presented in this paper can easily be extended to practical shaping and coding scenarios with non-zero rate loss. Fig. 1 illustrates how PCS realizes shaping gain and rate adaptation in our setup in the AWGN channel. At a fixed SNR, a lower-entropy PAM produces a larger NGMI than a higher-entropy PAM. As a result, PCS can pick the highest possible entropy that produces an NGMI greater than $NGMI^* = 0.8$ given the channel SNR. This allows the system to achieve the highest IR (cf. Eq. (3)) with error-free performance using the fixed-rate FEC. As shown in Fig. 1(inset), PCS PAM-8 signals achieve flexible rate adaptation with a fixed-rate FEC, and offer 0.15-bit/symbol gain in AIR (or 0.98-dB gain in SNR) with respect to the uniform PAM-4 signal. In the following sections, we will reveal various situations in which AIR gain and flexible rate adaptation may or may not be achieved, depending on the behavior of the transmitter.

III. LIMITATIONS OF IM-DD SYSTEMS

In this section, we briefly summarize the characteristics of IM-DD systems distinguished from coherent systems, which make essential difference in the design of PCS schemes. In our system modelling, $X$ denotes a generic transmitted signal at the baseband before optical modulation regardless of whether it is modulated on the optical amplitude or power. Similarly, $Y$ denotes a received baseband signal after optical demodulation.

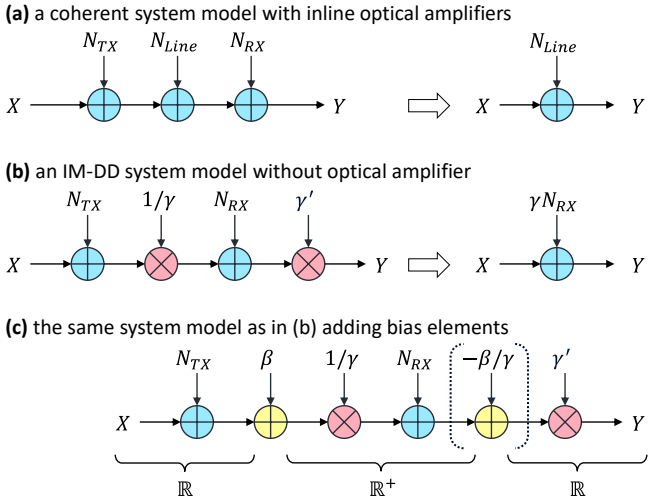

Fig. 2. The AWGN channel models for (a) a long-haul coherent system, (b) a short-reach IM-DD system without optical amplifiers, and (c) the same IM-DD system with biases $\beta$ and $-\beta/\gamma$ inserted.

Commonly, a coherent transmitter *linearly* transforms $X$ into the amplitudes of the optical field; while an intensity modulator *linearly* transforms $X$ into the optical *power*.

### A. Channel models for IM-DD systems

A coherent optical transmission system can be modeled as a concatenated AWGN channel [27]
$$Y = X + N_{TX} + N_{Line} + N_{RX} \tag{4}$$
as depicted in the left-hand side of Fig. 2(a), where $N_{TX}$, $N_{RX}$ and $N_{Line}$ are the transmitter, receiver, and optical link noises, respectively. The transmitted optical signal power is recovered in each fiber span by optical amplifiers, leading to a unit-gain channel. As a long-haul optical fiber link contains many inline amplifiers, the optical ASE noise overwhelms the transceiver noise, and predominantly determines the system performance. Therefore, the channel model in Eq. (4) can be degenerated into the form of
$$Y \approx X + N_{Line} \tag{5}$$
as depicted in the right-hand side of Fig. 2(a). $N_{Line}$ is usually treated as the white Gaussian noise and the system performance is evaluated in this *auxiliary* AWGN channel. In this simplified AWGN model, $N_{Line}$ includes both the linear and nonlinear impairments of the optical fiber channel, which is justified since the NLI can be modelled approximately as Gaussian noise under certain conditions [4].

In a channel modelling perspective, a critical difference of a short-reach IM-DD system from a long-haul coherent system is that the optical amplifier is *rarely* used in the link to maintain the low cost and power consumption. Correspondingly, the ASE noise no longer represents a predominant impairment, and the signal is subject to large attenuation at the receiver due to the lack of optical amplifiers. Note the absence of $N_{Line}$ and the inclusion of the multiplication by $1/\gamma \ll 1$ in the left-hand side of Fig. 2(b). The scaling factor $1/\gamma$ incorporates all the channel loss, *e.g.*, the insertion loss of external modulators and the fiber attenuation. The receiver then recovers the signal power at a scaling factor $\gamma' \gg 1$ which incorporates all the gain elements at the receiver, *e.g.*, a transimpedance amplifier (TIA) and a digital automatic gain controller (AGC). Without loss of generality, we use a single parameter $\gamma = \gamma'$ to model the end-to-end channel with unit gain, considering the digital scaling does not influence the receiver SNR. The IM-DD system can then be modelled as a concatenated AWGN channel
$$\begin{aligned} Y &= \gamma[(X + N_{TX})/\gamma + N_{RX}] \\ &= X + N_{TX} + \gamma N_{RX} \end{aligned} \tag{6}$$
as shown in the left-hand side of Fig. 2(b). Since $\gamma$ is a large number (*e.g.*, passive optical networks commonly require a link power budget >30 dB [28]), the system performance is almost exclusively determined by $N_{RX}$. Therefore, neglecting $N_{TX}$, the IM-DD system model in Eq. (6) can be approximated as
$$Y \approx X + \gamma N_{RX} \tag{7}$$
as depicted in the right-hand side of Fig. 2(b). $N_{RX}$ is commonly treated as white Gaussian because a DD receiver, consisting of a photodetector (PD) and an optional RF amplifier like a TIA, is mainly limited by the thermal noise.

Another distinct characteristic of the IM-DD system relative to the coherent system is that the transmitter modulates signals on optical *intensities* instead of optical fields. This causes confusion as if only the non-negative real numbers are allowed for the signal $X$ in Eq. (7), *i.e.*, as if $X \in \mathbb{R}^+$. In order to incorporate the fact that non-negative entities are transmitted through the optical fiber, we extend the system model in Fig. 2(b) as shown in Fig. 2(c). The signal $X$ is initially a symmetric bipolar one in the electrical domain, hence $X \in \mathbb{R}$. The intensity modulator then adds a bias $\beta$ to convert the signal to a unipolar one. Without loss of generality, we model the receiver as subtracting the bias from the unipolar signal after the optical-to-electrical (O/E) conversion. Though the bias subtraction may not be fulfilled by some receiver realizations, it has no influence on the performance analysis as the bias subtraction appears after all the noise elements in the model. If the constraint imposed on the system is appropriately quantified, whether it is an APC or an PPC, the shift elements at the transmitter and receiver can be removed from the system as they cancel out each other, and the model reduces back to Eq. (7) with bipolar signal $X$.

### B. Power constraint and channel quality metric

All the real-world electrical-to-optical (E/O) transmitters are fundamentally limited by a peak-power constraint (PPC)
$$max[|X|^2] \leq P_{peak}. \tag{8}$$
However, after E/O conversion, a coherent system typically uses an optical amplifier to boost the signal to yield a desired launch power hence the PPC is invalidated. The optical power is instead limited by NLI manifested during transmissions [3], and therefore a long-haul coherent system as modelled in Fig. 2(a) is subject to an APC
$$E[|X|^2] \leq P_{avg}. \tag{9}$$
In contrast, for the IM-DD system without optical amplifiers as modelled in Fig. 2(b), the PPC of Eq. (8) represents the only legitimate constraint imposed on the transmitted signals $X$.

If the system were subject to an APC, SNR would be a proper metric to characterize the channel quality. However, since the IM-DD system as in Eq. (7) is subject to a PPC, a proper channel metric would be the *peak-signal-to-noise ratio (PSNR)*

defined as
$$PSNR = \frac{max[|X|^2]}{\gamma^2 E[|N_{RX}|^2]}. \quad (10)$$
Consequently, under a PPC, the PSNR sensitivity represents a performance metric to evaluate different modulation schemes; namely, a modulation scheme is considered to offer a better receiver sensitivity if it requires a lower channel PSNR to meet an NGMI threshold. If the total channel loss $\gamma$ varies by $\Delta\gamma$ dB given the same $max[|X|^2]$ at the transmitter and the same $E[|N_{RX}|^2]$ at the receiver, it follows from Eq. (10) that
$$\Delta PSNR\ (dB) = -2 \cdot \Delta\gamma\ (dB). \quad (11)$$
This implies that the PSNR decreases by $2\Delta\gamma$ dB when the total channel loss $\gamma$ increases by $\Delta\gamma$ dB, in other words, the PSNR sensitivity is directly translated into a power budget. The fact that the PPC instead of the APC represents the proper constraint imposed on the IM-DD systems makes a critical difference to the PCS strategies, as will be described below.

*C. PAPR and signal quality metric*

While PSNR is the proper metric to characterize the IM-DD system's *channel* quality, it is the SNR given by
$$SNR = \frac{E[|X|^2]}{\gamma^2 E[|N_{RX}|^2]} \quad (12)$$
that evaluates the received *signal* quality. A crucial metric to analyze PCS for IM-DD systems in the following sections is the peak-to-average power ratio (PAPR) of $X$ defined as
$$PAPR = \frac{max[|X|^2]}{E[|X|^2]}. \quad (13)$$
The PAPR associates the PSNR and SNR as
$$PSNR = PAPR \cdot SNR \quad (14)$$
or equivalently,
$$SNR = \frac{PSNR}{PAPR}, \quad (15)$$
which is evident from Eqs. (10), (12), and (13).

An ideal bandwidth-unconstraint transmitter with an impulse response of delta function can generate a signal without distortion. In this case, the signal statistics are determined solely by the modulation format, and the peak power is given by the outmost modulation level. For example, an ideal non-return-to-zero (NRZ) signal has a PAPR of 0 dB. However, all real-world system components are band-limited, which inevitably induce distortions. Taking the NRZ signal again as an example, its eye diagram under the bandwidth constraint exhibits two thicker peak levels, which is usually referred to as peak distortion. Such distortion enhances the peak power, and hence, the PAPR. A severer bandwidth constraint brings higher PAPR enhancement (PE). Besides the intrinsic component bandwidth constraint, a digital transmitter can actively limit the signal bandwidth by digital pulse shaping like Nyquist shaping to accommodate the component bandwidth constraint. It may also employ various types of predistortion like pre-equalization and precoding to deal with the inter-symbol-interference (ISI) induced by the bandwidth constraint. All these factors can increase the degree to which the PAPR is enhanced.

According to Eq. (14-15), though the channel quality in the form of PSNR is fundamentally determined by the underlying physical system, the signal quality is determined jointly by the channel (*i.e.*, PSNR) and the signal's statistical property (*i.e.*, PAPR). As a result, PAPR plays a critical role in determining if a PCS design is optimal to maximize the shaping benefits. To this end, Section IV studies the PCS without transmitter PE assuming an ideal transmitter, and Section V extends the scope to the transmitter with PE that represents a realistic scenario.

### IV. PCS WITHOUT TRANSMITTER PE

In this section, we first analyze various PCS schemes in a PPC system when the transmitter induces no PE, in which case the signal PAPR is determined by the modulation format alone.

*A. PCS for Unipolar Signals*

The transmitted IM signal is treated as a non-negative signal in some literature [10]-[13], with a *unipolar* alphabet $\mathcal{X}^+ \coloneqq \mathcal{X} + \beta$, where $\mathcal{X} \coloneqq \{\pm 1, \pm 3, \ldots, \pm(M-1)\}$ is a bipolar PAM-$M$ alphabet symmetric around zero and $\beta \geq M - 1$ is the bias as shown in Fig. 2(c). Among all the distributions applicable on $\mathcal{X}^+$, an asymmetric MB (AS-MB) distribution maximizes the entropy $H(X)$ for the same average signal power $E[|X|^2]$, and equivalently, it minimizes $E[|X|^2]$ for the same $H(X)$. To see how energy efficiency of the AS-MB distribution is translated into better noise resiliency under an APC, we depict in Fig. 3(a) the relative positions and probabilities of AS-MB distributed signals drawn from the alphabet $\Delta\mathcal{X}^+$, where $\beta$ is set to $M - 1$ as an example (making the minimum intensity to zero) and $\Delta$ scales the IM signal to meet the APC. Clearly, the Euclidean distance between the adjacent intensity levels increases as $H(X)$ decreases, making the signal more tolerant to AWGN. As the APC is applicable to optically amplified IM transmission with either DD [10] or coherent detection [29],[30], the AS-MB distribution over a unipolar alphabet can achieve shaping gains.

However, without optical amplifiers in the system, an IM-DD system is subject not to an APC but to a PPC. Fig. 3(b) shows the AS-MB distributed signals $\Delta\mathcal{X}^+$ under a PPC, where $\Delta$ is constant regardless of $H(X)$. In contrast to the APC case of Fig. 3(a), change of $H(X)$ by PCS does not affect the Euclidean distance at all under a PPC [31]. This confirms the fact that it is not an MB distribution but the uniform distribution (cf. $H(X) = 3$ in Fig. 3(b)) that maximizes the entropy under a PPC, as mentioned in Section II.

To verify different impacts of PCS on PPC systems, we use Monto-Carlo simulations to evaluate the NGMI of PAM signals under various channel PSNR. Fig. 4(a) shows the NGMI of AS-MB distributed unipolar PAM-8 signals in the AWGN channel. At a given PSNR, the NGMI barely changes with $H(X)$ which coincides well with the analysis of Fig. 3(b), but changes significantly when the constellation size $M$ reduces from 8 to 4. From the perspective of PAPR, Fig. 3(c) explains why the reduction of $H(X)$ by PCS does not enhance the NGMI under a PPC. As $H(X)$ decreases, the *PAPR* (dashed line) increases at the same rate as the average power decreases; consequently, the *SNR* (solid line) decreases at the same rate as the PAPR increases according to Eq. (14). Thus, while reducing $H(X)$ at a fixed SNR brings a PCS gain in a coherent system, in an IM-DD system having a fixed PSNR, reducing $H(X)$ accompanies reduction in *SNR* and hence the PCS gain is cancelled out.

It is worth noting that the above analysis does not match what was published in early literatures [11]-[13] where benefits from unipolar PCS schemes are demonstrated in IM-DD systems without optical amplifiers. This discrepancy arises from a different channel metric, namely, the received optical power (ROP) in those literatures instead of the PSNR in this work. The ROP represents an equivalent metric of PSNR for the channel conveying a bipolar signal symmetrically distributed around $\beta$, whose optical power is determined solely by the modulator bias $\beta$ and is independent from the signal's intensity distribution, as shown in Fig. 3(d). On the other hand, if PCS produces an intensity distribution that is *not* symmetric around $\beta$, the optical power is altered by the distribution, as shown in Fig. 3(d) for the unipolar signal with AS-MB distributions. In this case, the optical power at the receiver decreases as $H(X)$ reduces, which can be translated to higher receiver sensitivity. However, the signal loses the same amount of power at the transmitter, and the channel power budget (*i.e.*, $1/\gamma$) remains unchanged. In such a sense, the PSNR is a more proper metric than the ROP in PPC systems, as it straightforward characterizes the power budget as in Eq. (10). Despite this, the findings of the AS-MB distributed unipolar signals [11]-[13] are still meaningful and can be applied to the amplified IM-DD system with an optical APC [10,29,30]. For example, the pairwise MB distribution [11,12] remains the most practical scheme for unipolar PCS to be compatible with the PAS architecture.

*B. PCS for Bipolar Signals*

Based on the discussion in Section IV-A, we treat the signal in an optical-amplification-less PPC system as a *bipolar* signal and evaluate PCS schemes for a bipolar PAM alphabet. In this case, symmetric MB distributions cannot offer better energy efficiency than uniform distributions under a PPC, for the same reasons discussed with Fig. 3(b); namely, under a PPC, PCS does not change the minimum Euclidean distance of the constellation, and hence the tolerance to AWGN. As expected, Fig. 4(b) shows that the NGMI of MB-distributed signals barely changes with $H(X)$ at a fixed PSNR.

Another symmetric distribution of interest for a PPC system is the *reverse* MB (R-MB) distribution. R-MB distributions are formed by removing the negative signs before the parameters $\lambda$ ($\lambda \geq 0$) in Eq. (2). In contrast to the MB distribution, the R-MB distribution assigns higher probabilities of occurrence for the outer levels of a bipolar alphabet than for the inner levels. While an MB distribution *minimizes* the average signal energy for a given $H(X)$, an R-MB distribution *maximizes* the average signal energy. The rationale behind the R-MB distribution in PPC systems is that it maximizes the SNR for the same noise power. Such SNR enhancement cannot improve the tolerance to AWGN, again due to the unchangeable Euclidean distance. In essence, under the same $H(X)$, an R-MB distributed signal intrinsically requires higher SNR than the MB distributed signal to achieve the same NGMI target. However, we do observe in Fig. 4(c) that the NGMI of R-MB distributed signals gradually increases as $H(X)$ decreases. The unexpected phenomenon is attributed to a unique characteristic of R-MB distributions that produce most frequently the outermost two levels that have only one neighbor placed at the minimum Euclidean distance unlike the inner levels having two neighbors. This decreases erroneous symbol decisions as the error probability of the outermost levels is smaller than that of the inner levels. The shaping gain in Ref. [17] achieved by R-MB-like distributions may be explained in this respect. Note that such a gain comes from a different cause than what the classical PCS has been designed for, and have two distinct features from the MB-distributed PCS: *(i)* the gain decreases as the PAM order increases since the relative fraction of the two outermost levels reduces, and *(ii)* the rate adaptability deteriorates as the PSNR increases (*cf.* Fig. 4(c)) and almost vanishes in the high NGMI regime where low-overhead FEC typical for IM-DD systems are applicable.

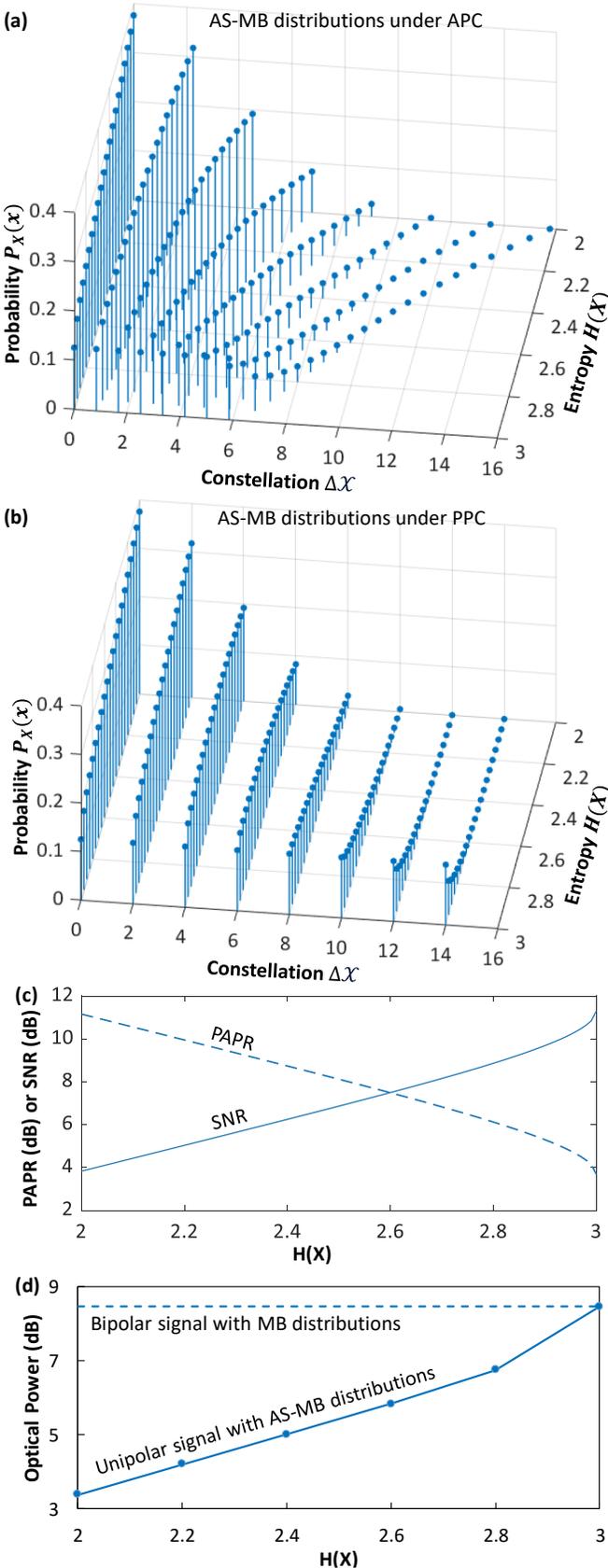

Fig. 3. AS-MB distributions of PCS PAM-8 signals over a unipolar alphabet $\Delta\mathcal{X}^+$ under the (a) APC (b) PPC; (c) PAPR (dashed line) for unipolar PAM-8 signals as a function of $H(X)$ and the corresponding SNR (solid line) when $PSNR = 15$ dB; (d) optical signal power as a function of $H(X)$, when signals follow MB distributions (dashed line) and AS-MB distributions (solid line), on the alphabet of $\mathcal{X}^+ = \mathcal{X} + \beta$ where $\mathcal{X} = \{\pm 1, \pm 3, \pm 5, \pm 7\}$ and $\beta = 7$.

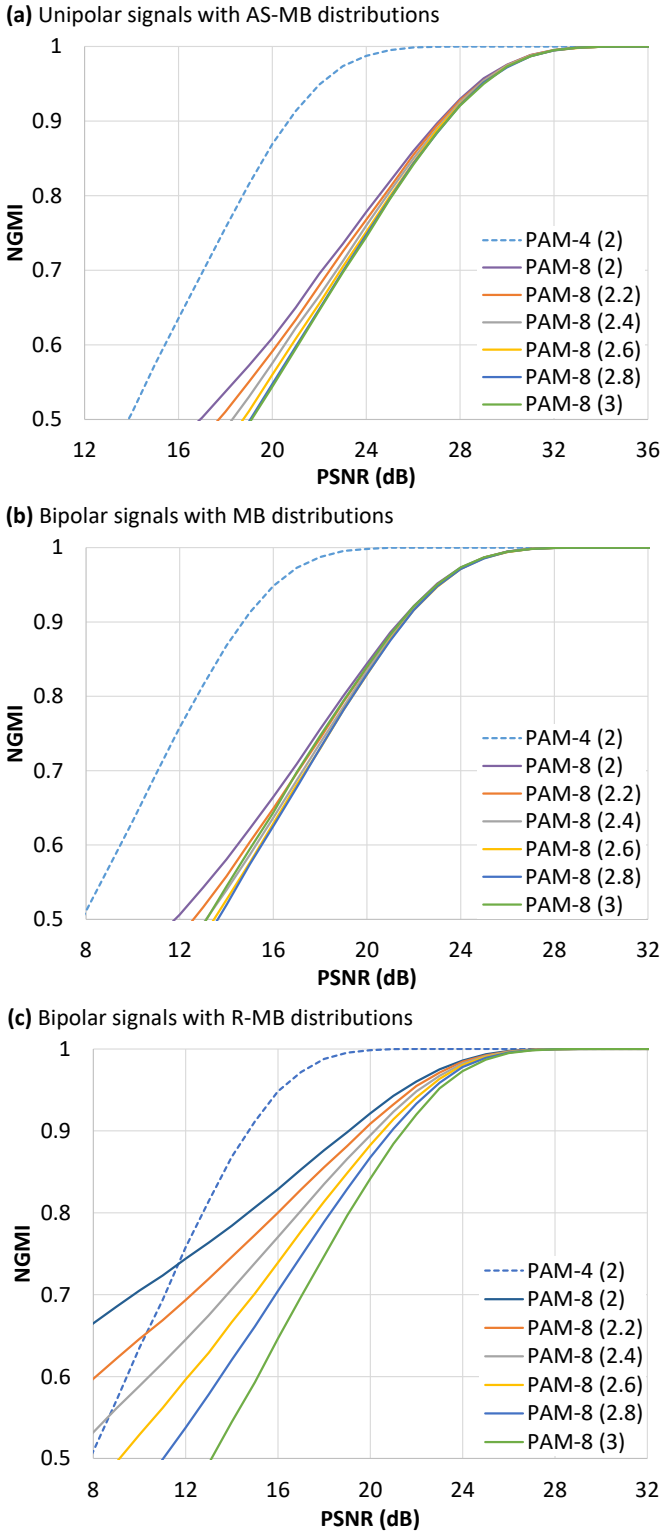

Fig. 4. NGMI in PPC systems: (a) unipolar signals with AS-MB distributions; bipolar signals with (b) MB distributions and (c) R-MB distributions.

## V. PCS WITH TRANSMITTER PE

As mentioned in Section III.C, all components inside a real-world transmitter have intrinsic bandwidth constraints that can enhance the signal PAPR. As revealed by Eqns. (14-15), the PE at the transmitter reduces the SNR given a fixed channel PSNR. Therefore, the transmitter PE should be taken into account when analyzing PCS schemes. In this section, we consider Nyquist pulse shaping to study the influence of PE on PCS performance. Though Nyquist pulse shaping is chosen for an illustration purpose, any PE effects from the transmitter can be analyzed by the same approach. Namely, the PAPR can fully characterize the quantitative relation between the channel PSNR and PCS performance regardless of what causes the PE.

### A. Transmitter PE by Pulse Shaping

We focus on an IM-DD system with a pair of root-raised cosine (RRC) filters $g(\cdot)$ at the transmitter and the receiver, respectively. Fig. 5 shows the probability density functions (PDFs, left figures) and complementary cumulative density functions (CCDFs, right figures, to be described below in more detail) of a PAM-8 signal $X$ (green solid lines), as well as its RRC-filtered signal $g(X)$ with roll-off factors of $\rho = 0.4$ (blue dashed lines) and $\rho = 0.01$ (orange solid lines). $X$ follows a uniform distribution in Fig. 5(a) and an MB-distribution with $H = 2.2$ in Fig. 5(b). The PDF and CCDF of the zero-mean Gaussian distribution with a variance of $E[|X|^2]$ is also shown (black dotted lines) as a reference. Clearly, the RRC filtering at the transmitter transforms a discrete signal towards a Gaussian distribution, as the roll-off factor $\rho$ decreases.

By comparing Fig. 5(a) and Fig. 5(b), we notice that the peak power gap between the uniform- and MB- distributed signals is significantly shrunk with a small RRC roll-off factor of $\rho = 0.01$. Since the peak power tends towards infinity as the distribution of the signal approaches a Gaussian distribution, we use a widely adopted metric to quantify the PAPR of signals with an arbitrary distribution [32], where the peak signal value is determined by a limit $\sigma^2$ such that a fraction $\varepsilon$ (*i.e.*, the clipping ratio) of the signal has a power greater than $\sigma^2$. Formally, the CCDF

$$CCDF(x) = \Pr(|X|^2 \geq x) \qquad (16)$$

quantifies the probability of the signal power being greater than $x$ (cf. right-hand side figures of Fig. 5), which can be obtained by integrating the PDF of $X$ (cf. left-hand side figures of Fig. 5) from $x$ to $\infty$. Then, the PAPR is conveniently written by

$$PAPR(\varepsilon) = \frac{\sigma^2}{E[|X|^2]} \qquad (17)$$

with $\sigma^2$ being determined such that $CCDF(\sigma^2) = \varepsilon$. Indeed, this $PAPR(\varepsilon)$ represents the PAPR of a true physical system, since no physical system can transmit unbounded $X$ without clipping it at a finite limit $x$. In what follows, we choose $\varepsilon = 10^{-5}$ to quantify the PAPR.

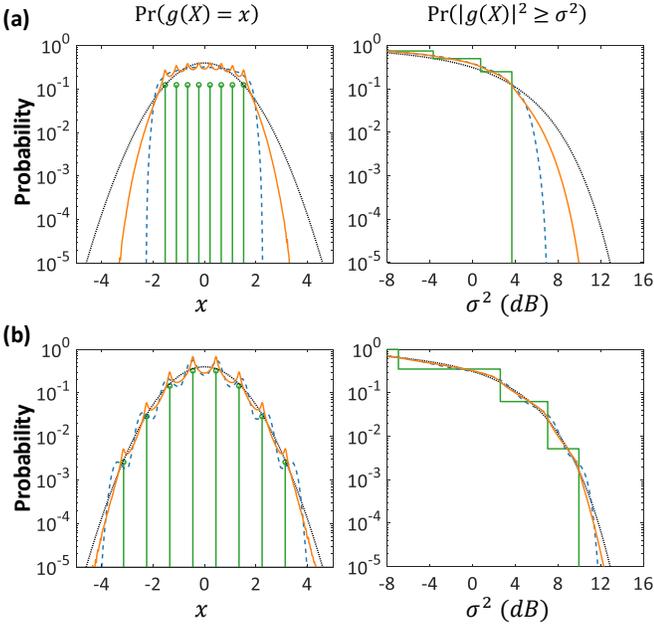

Fig. 5. PDF of $g(X)$ (left) and CCDF of $|g(X)|^2$ (in $dB$) (right) for PAM-8 signals with (a) uniform- and (b) MB- distribution ($H(X) = 2.2$), before RRC filtering (green lines) and after RRC filtering with roll-off factors of $\rho = 0.4$ (blue dashed lines) and 0.01 (orange solid lines). For all the curves, the signal power is normalized to 1. The Gaussian distribution $N{\sim}(0,1)$ is also shown (black dotted lines) as the reference.

Before RRC pulse shaping, the PAPR of the uniform PAM-8 signal is 6.3 dB lower than that of the MB-distributed PAM-8 signal ($H = 2.2$); however, after RRC filtering with $\rho = 0.01$, the PAPR of the uniform PAM-8 signal is only 2.4 dB lower than that of the MB-distributed signal. Apparently, while the PCS scheme with MB distribution enhances the PAPR, such an enhancement effect has been weakened by RRC filtering.

### B. Quantitative relation between PE and PCS performance

In order to see the impact of PE on PCS in a PPC system, we first investigate how the PAPR changes the minimum Euclidean distance of a PAM signal. Fig. 6 shows the relative positions and probabilities of MB-distributed signals in alphabet $\Delta \mathcal{X}$ which are RRC-filtered with $\rho = 0.4$ (cf. Fig. 6(a)) and 0.01 (cf. Fig. 6(b)). Fixed the peak limit, all the signals in Fig. 6 are made to be clipped with the probability of $\varepsilon = 10^{-5}$ after RRC filtering, by adjusting the constellation scaling factor $\Delta$ for each entropy $H(X)$. For the large roll-off factor $\rho = 0.4$ in Fig. 6(a), the peak-power enhancement due to RRC filtering is insignificant for both the uniform (cf. $H(X) = 3$ in Fig. 6) and MB-distributed signals, and the Euclidean distance does not alter much with varying $H(X)$ (as similarly observed in the absence of RRC filtering in Fig. 3(b)). However, for the small $\rho = 0.01$ in Fig. 6(b), the peak-power enhancement due to RRC filtering is much more noticeable for the uniform signal than for the MB-distributed signals, leading to a much smaller Euclidean distance for the uniform signal to meet the same PPC than for the MB-distributed signals. As a result, the transmitter PE relaxes the PPC and converts it towards an APC by transforming discrete signals with arbitrary distributions to continuous signals with Gaussian-like distributions.

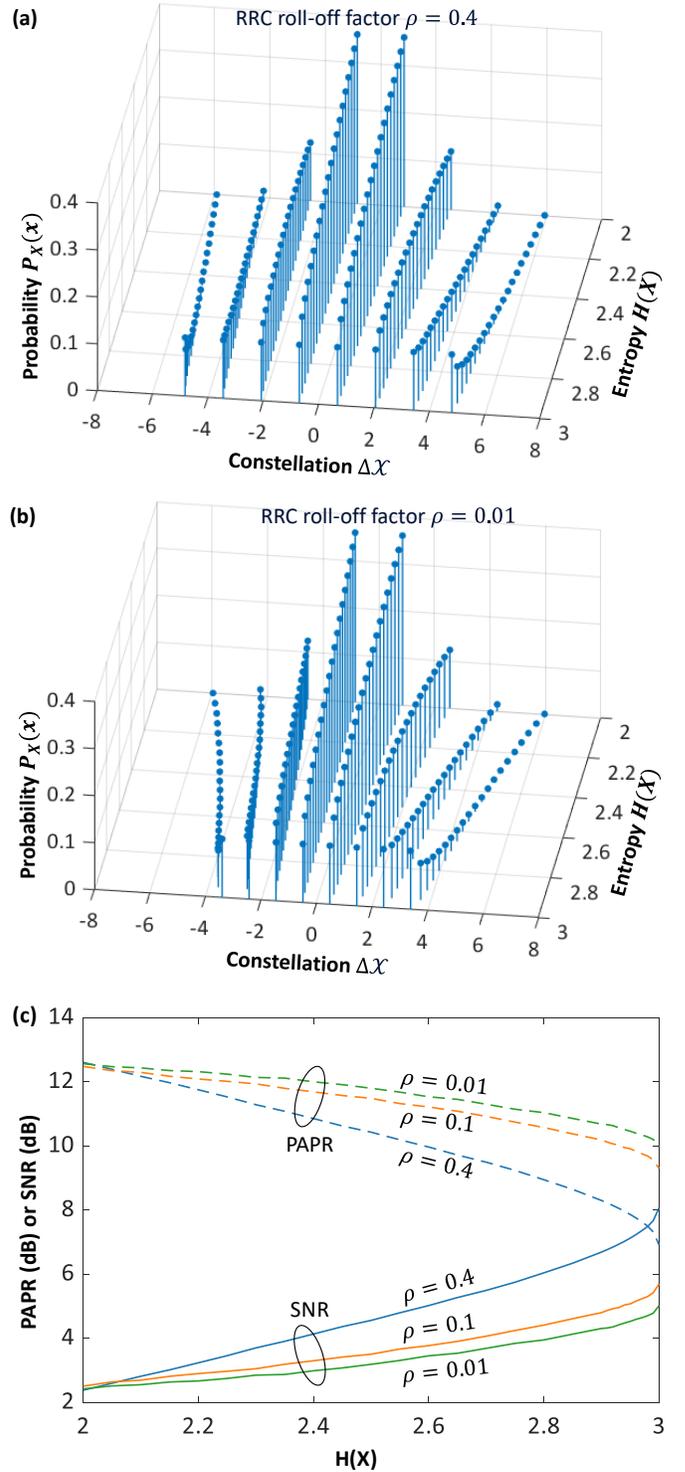

Fig. 6. MB-distributions of PCS PAM-8 signals over a bipolar alphabet $\Delta \mathcal{X}$ when RRC filtering is performed with (a) $\rho = 0.4$ and (b) $\rho = 0.01$, under a PPC: $max[|g(X)|^2] \leq 49$; (c): PAPR (dashed lines) of MB-distributed signals as a function of $H(X)$ after RRC filtering, and the corresponding $SNR$ (solid lines) when $PSNR = 15$ dB.

As revealed by Eqs. (14)-(15), when a fixed PSNR is given by the IM-DD channel, the SNR is inversely proportional to the PAPR. Fig. 6(c) shows the SNR and PAPR curves for MB-distributed PAM-8 signals given the PSNR of 15 dB. As the PCS shapes harder on the PAM alphabet (*i.e.*, as it decreases $H(X)$), the PAPR increases and the SNR decreases in general; and the RRC roll-off factor $\rho$ determines how much the PAPR (and consequently the SNR) changes with $H(X)$. When $\rho$ is large (see $\rho = 0.4$ in Fig. 6(c)), the PAPR increases and the SNR decreases at a fast rate when PCS reduces $H(X)$, hence offsetting the increased Euclidean distance by PCS for the same average signal power, similarly to the case without transmitter PE as in Fig. 3(c). However, with smaller RRC roll-off factors, the influence of PCS on the PAPR is weakened and the SNR changes much less with $H(X)$. In the case when the transmitter induces significant PE making the PAPR approach a constant over all shaping factors (we refer to it as "extreme PE" in what follows), the SNR approaches a constant regardless of the PCS shaping factors. In this case, a PPC system is converted to an APC system where the benefits of PCS can be fully exploited.

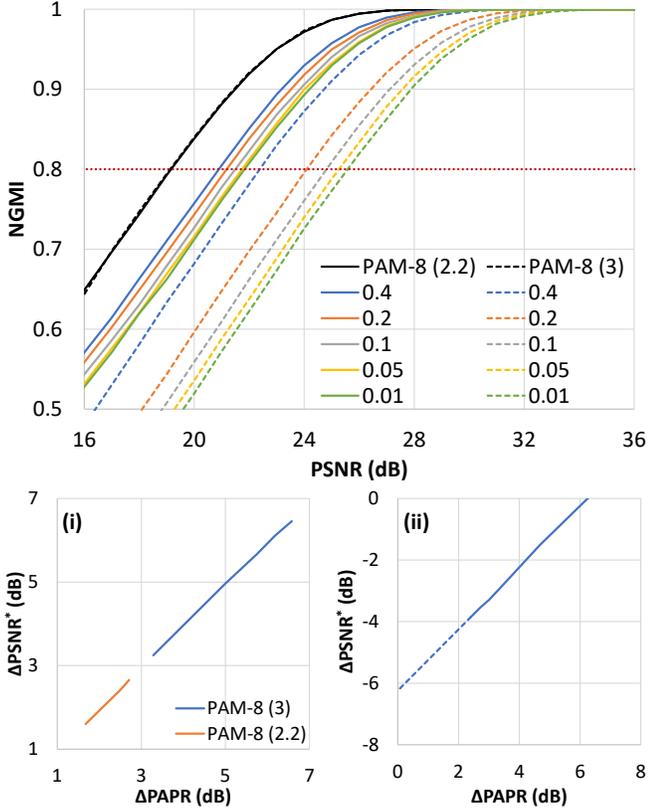

Fig. 7. Comparisons of NGMI between the uniform PAM-8 signal and the MB-distributed PAM-8 signal ($H = 2.2$). The numbers in the legend are the roll-off factors of the RRC filter. Insets show the relation between $\Delta PAPR$ and $\Delta PSNR$ keeping (i) the same modulation, (ii) the same RRC roll-off factor.

Most IM transmitters would be operated in an intermediate condition between the extreme-PE condition as analyzed above and the no-PE condition as analyzed in Section IV. In such an intermediate condition, the PCS gain under a PPC can still be quantitively characterized by using the PAPR; namely, it follows from Eqs. (14)-(15) that if a modulated signal has a larger PAPR than a chosen reference signal by $\Delta PAPR \ [dB] \coloneqq PAPR[dB] - PAPR_{Ref}[dB]$, it requires a larger PSNR than the reference signal by

$$\Delta PSNR^* = \Delta PAPR + \Delta SNR^* \text{ (all in } dB\text{)} \quad (18)$$

to achieve the same NGMI, where $\Delta PSNR^*[dB] \coloneqq PSNR^*[dB] - PSNR^*_{Ref}[dB]$ with $PSNR^*$ being the PSNR threshold to achieve the NGMI threshold $NGMI^*$, and $\Delta SNR^*[dB] \coloneqq SNR^*[dB] - SNR^*_{Ref}[dB]$ with $SNR^*$ being the SNR threshold to achieve $NGMI^*$ in the AWGN channel. We evaluate the system's NGMI with various RRC roll-off factors for two modulation formats, the uniform PAM-8 and the MB-distributed PAM-8 with $H(X) = 2.2$, as shown in Fig. 7. As expected, the NGMI curves of the PCS and the uniform signals overlap with each other when the transmitter induces no PE. As the RRC roll-off factor decreases, the NGMI of the uniform PAM decreases substantially while that of the PCS PAM remains similar, hence the PCS gain is better manifested. We set $NGMI^* = 0.8$ to evaluate the system's PSNR sensitivity. For each modulation (*i.e.*, uniform PAM-8 *or* MB-distributed PAM-8 with $H(X) = 2.2$), we compare among various RRC roll-off factors by setting the signal without transmitter PE as the reference. As $\Delta SNR^* = 0$ in Eq. (18) when comparing the same modulation, $\Delta PSNR^* = \Delta PAPR$. This is clearly shown in Fig. 7(i). Further, for each pair of uniform and MB-distributed ($H(X) = 2.2$) signals with identical RRC roll-off factor, we compare among various roll-off factors by setting the uniform signals as the reference. Again, $\Delta PSNR^*$ increases linearly with $\Delta PAPR$, as shown in Fig. 7(ii). For the extreme-PE case when $\Delta PAPR = 0$, a PCS signal achieves the PSNR gain of 6.2 dB over the uniform signal (*i.e.*, $\Delta PSNR^* = \Delta SNR^* = 6.2$ in Eq. (18)), the same as in the APC system shown by Fig. 1.

### C. Revisiting R-MB PCS

As presented in Section IV-B, without transmitter PE, R-MB-distributed signals show some gain over the uniform signal at the low PSNR region. When the transmitter PE exists, the PAPR gap between the R-MB-distributed and the uniform signals shrinks for the same reason as described for the MB-distributed signal in Section V-A. However, in the absence of the transmitter PE, the PAPR of R-MB distributed signals is lower than that of the uniform signal, whereas the PAPR of the MB-distributed signals is higher than that of the uniform signal. Therefore, the narrowed PAPR gap reduces the gain for PCS signals with R-MB-like distributions.

### D. Discussions on the extreme-PE condition

Since the significance of time-domain coefficients of finite impulse response (FIR) RRC filters gradually vanishes towards both ends of the filter, RRC filters cannot produce a PAPR like the ideal Gaussian signal that has the PAPR independent from the modulation format (*i.e.*, the extreme PE condition), even if the roll-off factor $\rho \to 0$. Nevertheless, the PAPR dependence on the modulation format can be further weakened with the combination of a number of transmitter PE effects, including the digital pulse shaping analyzed in Section V.A, digital pre-distortion like pre-equalization and pre-coding, component bandwidth constraint, and so on. The impact of these joint PE effects requires case-by-case analysis and is out of the scope of this paper. In general, once the PAPR is evaluated after applying all the PE effects, the PCS performance can be predicted in the same way as described in Section V.B, regardless of the various reasons behind the PE.

Notably, there exists a widely-used modulation technique of particular interest that makes arbitrary modulations follow the Gaussian distribution in time domain, namely, the multicarrier (MC) modulation. The PAPR of MC signals is usually sensitive to the number of subcarriers and the oversampling ratio [32], but its dependence on the modulation format is very weak. In other words, MC produces almost the same PAPR regardless of the modulation format of subcarriers, hence converts the system constraint from a PPC to an APC. When the subcarriers are filled with rate-adaptive symbols that have the exact entropy matching the underlying channel condition, the modulation is commonly referred to as entropy loading [20]. As a PPC is almost equivalent to an APC in most multicarrier systems that resembles the extreme-PE condition, entropy loading can take advantages of PCS in PPC systems. This explains the gain observed in previous entropy loading IM-DD systems [20],[21].

While the extreme-PE condition can take the full advantage of PCS benefits in PPC systems, the design of an optimal IM format aiming at a minimum cost per bit should involve various considerations besides the PCS like the component cost, DSP complexity, power consumption and so on. Note that although the rate adaptability is degraded in the moderate PE case, it may be useful in some applications.

## VI. Experiment

In this section, we focus on the MB-distributed PCS scheme and investigate its benefit in a state-of-the-art IM-DD system with different amount of transmitter PE. The system setup is shown in Fig. 8. We generate PAM signals at variable symbol rates by an 8-bit 120-GSa/s digital-to-analog converter (DAC), which are then amplified by a 55-GHz RF amplifier. The RF signals drive an O-band external modulated laser (EML) module consisting of a distributed-feedback (DFB) laser and an electro-absorption modulator (EAM). The EML has a 3-dB bandwidth of 32 GHz and a 6-dB bandwidth of 40 GHz. The signals are tested in the back-to-back configuration, with an optical attenuator inserted in front of the receiver for sensitivity measurement. The signal is detected using a 70-GHz PIN photodiode (PD). Due to the lack of a transimpedance amplifier, the signal is amplified by a 60-GHz RF amplifier after the PD, and then digitized by a 63-GHz real-time oscilloscope sampling at 160-GSa/s. The receiver DSP resamples the signal at 2 samples per symbol (sps), and then performs matched filtering, timing recovery and 1000-tap least-mean-square equalization.

We investigate the benefit of PCS by varying the symbol rate of PAM signals to emulate three exemplary scenarios with different bandwidth limitations, as summarized in Table I. The first scenario emulates signals with no PE analyzed in Section IV. We choose a symbol rate of 40 GBaud and use a raised cosine (RC) filter with a roll-off factor of 1, whose bandwidth is limited within the 6-dB bandwidth of the EML. Such an RC filter does not induce PE after digital oversampling. In opposite, the third scenario emulates signals with "extreme PE" analyzed in Section V-B. We choose a symbol rate of 80 GBaud whose excess bandwidth is sharply cut off by an RRC filter with a roll-off factor of 0.05 to fully exhaust the transmitter bandwidth. Also, in order to compensate for the weak response of the DAC and the EML at high frequencies, we apply a frequency pre-emphasis filter that has 8 dB greater gain at the spectral edges than at the zero frequency. Even though the DSP combining the RRC and pre-emphasis filters induces significant PE, a PAPR gap of around 1 dB remains between the PCS and uniform signals. To accurately emulate the extreme-PE condition for scenario 3, we slightly scale all the 80-GBaud signals to have the same average power such that the PPC becomes equivalent to an APC. The second scenario is an intermediate case between the above two. We choose a symbol rate of 60 GBaud with a moderate roll-off factor of 0.2 for the RRC filter. The PAPR values without and with PE are compared in the inset of Fig. 9(b). Clearly, the PAPR difference between the uniform and PCS signals is greatly shrunk with PE, but is not ignorable.

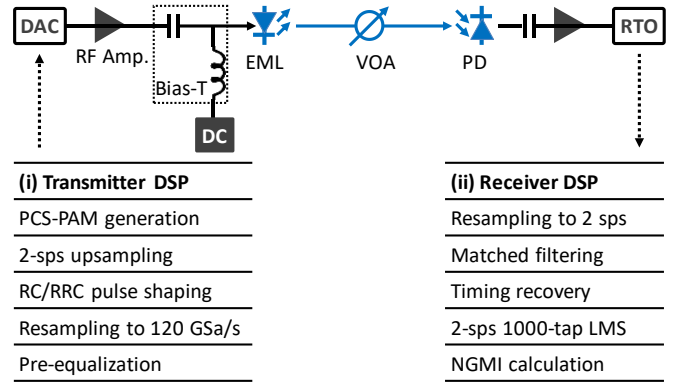

Fig. 8. Experiment setup. DAC: digital-to-analog converter; EML: external modulated laser; VOA: variable optical attenuator; PD: photodiode, RTO: real-time oscilloscope.

TABLE I
SUMMARY OF THE THREE EXPERIMENT SCENARIOS

| Scenario | 1 | 2 | 3 |
|---|---|---|---|
| Emulated objective | Minimum PE (Section IV-B) | Moderate PE (Section V-B) | Extreme PE (Section V-B) |
| Symbol rate | 40 | 60 | 80 |
| Nyquist filter | RC ($\rho = 1$) | RRC ($\rho = 0.2$) | RRC ($\rho = 0.05$) |
| Pre-equalization | None | Weak | Strong |
| Average power | Very different | Slightly different | Same |
| Rate adaptation under the PPC | Little | Moderate | Similar to the APC system |
| Shaping gain | No | No | Yes |

The measured NGMI at varying ROP is illustrated in Fig. 9(a-c) for all the three scenarios, and Fig. 9(d) compares the AIRs of PAM-8 with various entropies at the NGMI threshold of 0.8. For scenario 1, the NGMI hardly changes with $H(X)$. This implies PCS offers no rate adaptation in optical-amplifier-less IM-DD systems with minimum transmitter PE. In contrast, when the transmitter satisfies the extreme-PE condition in scenario 3, PCS achieves the shaping gain and rate adaptation range as in an APC system. For scenario 2, while PCS shows no shaping gain over the uniform PAM-4 signal, it achieves a moderate range of rate adaptation. Moreover, such a range can be quantitatively related to the PAPR values in Fig. 9(b). Taking the difference between uniform PAM-8 ($H = 3$) and PCS PAM-8 ($H = 2.2$) signals as an example, the rate adaptation range ($\Delta PSNR^*$) is 6.2 dB when $\Delta PAPR = 0$ (i.e., the extreme-PE case) as shown in Fig. 7(ii). The $\Delta PAPR$ of 2.9 dB after the moderate PE should shrink $\Delta PSNR^*$ to 3.3 dB that corresponds to an ROP adaptation of 1.7 dB according to Eq. (11). This is close to the experimental result of 2.0 dB as shown in Fig. 9(d). The narrower range predicted by the PAPR may be attributed to a factor smaller than 2 in Eq. (11) between $\Delta PSNR$ and $\Delta \gamma$ if the receiver noise is signal-power dependent, as is the case for this receiver setup with an RF amplifier that may produce higher nonlinear noise given higher input power.

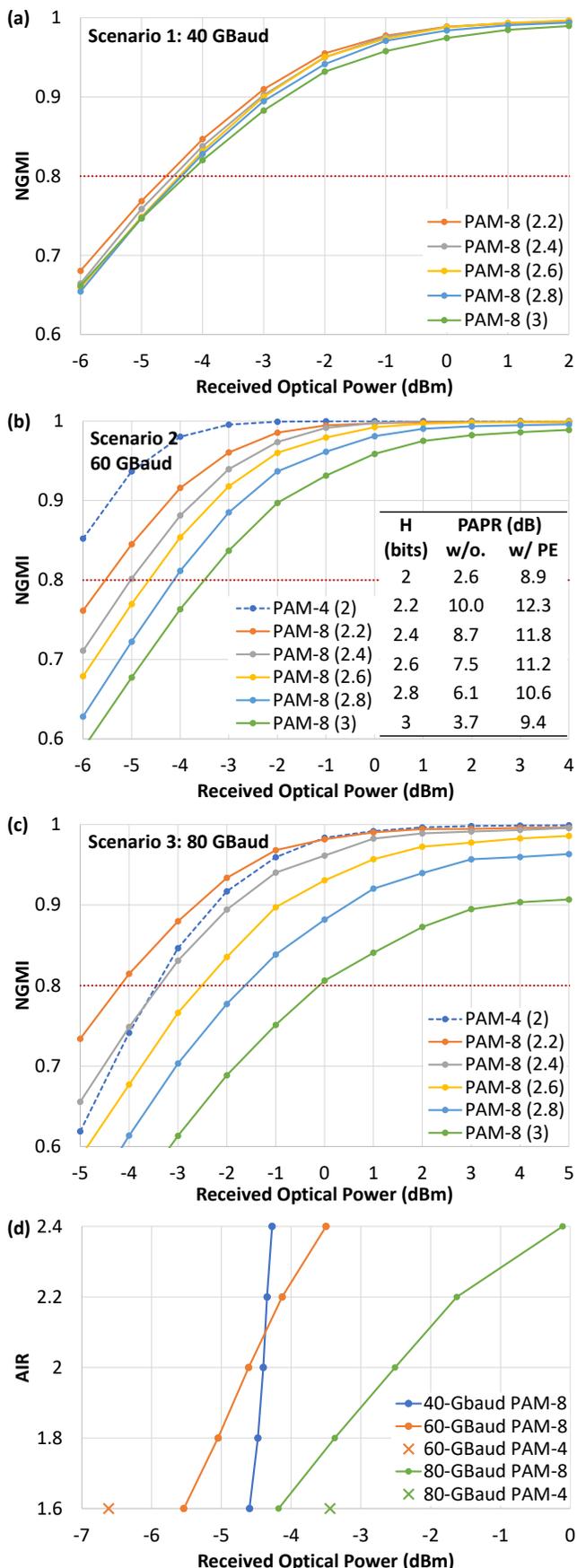

Fig. 9. (a-c) Experiment results of the three scenarios summarized in Table I; (d) comparisons of the rate adaptation range among the three scenarios.

## VII. CONCLUSION

We present a mathematical model for the IM-DD system that is subject to a PPC and study the benefits of various PCS schemes on the system. If an ideal transmitter induces little enhancement on PAPR, the traditional MB-distributed PCS cannot exploit the energy efficiency presented in the APC system, and neither the asymmetric MB distribution on a unipolar PAM alphabet nor the symmetric MB distribution on a bipolar alphabet can bring benefits to the IM-DD system. Some uncommon schemes like the R-MB distributed PCS may improve the symbol decision performance, despite the gain is attributed to a cause different from what the classical PCS has been designed for. A real-world bandwidth-limited transmitter inevitably induces PAPR enhancement due to a variety of factors like component bandwidth limitation, digital pulse shaping and pre-distortion. In this case, the PPC can be relaxed by the narrower PAPR gap between the PCS and uniform signals, and the traditional MB-distributed PCS can achieve some shaping benefits like the rate adaptation with a fractional range under the fixed-rate FEC. The PCS performance for a system with transmitter PE can be fully characterized by the signal PAPR regardless of what causes the PE. If the PAPR becomes independent from PCS modulations, for example, in a discrete multitone (DMT) system, the PPC and APC would be equivalent, and PCS can take full advantage of its shaping benefits.

ACKNOWLEDGMENT

We would like to thank *Source Photonics Inc* (West Hills, CA 91304, USA) for providing the EML in the experiment.